\documentclass{eptcs}
\providecommand{\event}{QPL 2026}

\usepackage{amsmath,amssymb,amsfonts}
\usepackage{mathtools}
\usepackage{hyperref}

\usepackage[T1]{fontenc}
\usepackage[utf8]{inputenc}
\usepackage[english]{babel}

\usepackage{lmodern}

\usepackage{microtype}

\usepackage{amsmath,amssymb,amsthm,mathtools}

\usepackage{xcolor}   
\usepackage{enumitem} 

\theoremstyle{definition}
\newtheorem{definition}{Definition}
\newtheorem{lemma}{Lemma}
\newtheorem{proposition}{Proposition}

\newtheorem{theorem}{Theorem}
\newtheorem{observation}{Observation}

\usepackage{tikz}
\usetikzlibrary{trees,arrows.meta,positioning}
\usepackage{braket}

\newcommand{\Dom}{\mathrm{Dom}}

\newcommand{\NonHam}{\ensuremath{\mathrm{NonHam}}}
\newcommand{\Sub}{\mathrm{Sub}}
\newcommand{\HC}{\mathrm{HC}}
\newcommand{\imply}{\ensuremath{\supset}}
\newcommand{\WF}{\mathrm{WF}}
\newcommand{\Closed}{\mathrm{Closed}}
\newcommand{\lab}{\mathrm{label}}
\newcommand{\poly}{\mathrm{poly}}

\title{Semantic Entropy of Compressed Proofs:
Sheaf Geometry and Quantum Verification of DLDS Certificates}

\author{Lorenzo Saraiva \institute{Departamento de Informática PUC--Rio, Brazil}
\email{lsaraiva@inf.puc-rio.br}
\and Edward Hermann Haeusler
\institute{Departamento de Informática PUC--Rio, Brazil}
\email{haeusler@inf.puc-rio.br}
 \and Guilherme Lima \institute{Instituto PUC-Behring de IA, PUC--Rio, Brazil}
\email{glima@inf.puc-rio.br}
}

\def\titlerunning{Semantic Entropy of Compressed Proofs}
\def\authorrunning{L.\ Saraiva, E.\ H.\ Haeusler \& G.\ Lima}

\begin{document}
\maketitle

\begin{abstract}
We introduce \emph{semantic entropy}, a structural invariant of compressed
Dag--Like Derivability Structures (DLDS), and show that it governs both the
robustness and the quantum detectability of invalid proof certificates.
Assuming natural structural properties of compressed derivations, semantic
entropy equals the logarithmic multiplicity of global sections of a sheaf of
admissible local proof choices and simultaneously the minimal Hilbert-space
dimension hidden from closure-verification observables.

Polynomial semantic entropy implies inverse-polynomial density of invalid
certificates and efficient detection via quantum amplitude amplification.
Conceptually, this establishes a bridge between proof compression,
sheaf-theoretic geometry of information, and quantum verification complexity.
\end{abstract}

\section{Introduction}

The class \textbf{coNP} contains natural problems, such as graph
non-Hamiltonicity, tautology, and unsatisfiability, for which no short
classical certificates of membership are known. A central question in
proof complexity asks whether such certificates exist in some structured
form, and whether they can be efficiently verified. Aaronson and Kuperberg
proposed the class \textbf{QCMA}~\cite{aaronson2007quantum} as a quantum
analogue of NP in which the verifier is a polynomial-time quantum circuit.
The containment $\mathrm{coNP} \subseteq \mathrm{QCMA}$ is not known to
hold unconditionally, and identifying explicit polynomial-size
certificates for coNP-complete problems verifiable by a quantum machine
remains open.

Dag--Like Derivability Structures (DLDS), obtained from natural-deduction
proofs by horizontal compression, are candidate certificates of this kind.
Compression identifies repeated proof occurrences at the same depth,
shrinking exponential-size tree-like derivations to structures whose
kernel is polynomial in the size of the conclusion formula. The price is
that compression makes the discharge relation ambiguous: a single
open-assumption occurrence in the DLDS may have several candidate
$(\imply\text{-I})$ discharge points, only one of which is correct in any
given tree-like unfolding. A \emph{certificate} therefore consists of the
DLDS together with a \emph{selector string} resolving these ambiguities,
and verification reduces to checking a polynomial-time Boolean predicate
$f_{D}(x)$ over an exponentially large certificate space.

This is where quantum verification enters the picture. Classical
brute-force search over the $2^{m^{2}\log m}$ candidate selector
assignments is infeasible. The pipeline becomes polynomial only if
invalid certificates are sufficiently dense in the search space to be
detectable by amplitude amplification. The central question of this paper
is therefore: \emph{what structural property of compressed DLDS forces
inverse-polynomial density of counterexamples?}

We answer this by introducing a single invariant, \emph{semantic
entropy}, defined as the logarithm of the joint admissible-domain size
of the free selectors with respect to a closure-failure witness. Our main
result (Theorem~\ref{thm:entropy-growth}) shows that under two natural
structural assumptions on the implicational encoding of
non-Hamiltonicity, collapse-propagation (CP) and naive frontier
largeness (NF), semantic entropy grows as
$\Omega(m^{2} \log m / \poly(n))$. This is large enough to force
inverse-polynomial counterexample density
(Theorem~\ref{thm:geom-robustness-quant}) and hence polynomial-time
quantum detectability via BBHT~\cite{Brassard2002}.

A few design choices warrant explanation upfront. We work in the
\emph{implicational fragment of minimal logic} because it is the smallest
setting in which the encoding goes through: conjunction, disjunction, and
absurdity admit standard implicational simulations with polynomial
overhead~\cite{subformula}, accounting for the cubic bound
$m = |\NonHam(G)| = O(n^{3})$. We use \emph{non-Hamiltonicity} as the
running coNP-complete problem because its tree-like proofs have been
characterized in prior
work~\cite{haeusler2025horizontalcompressiondagderivationsminimal}; the
analysis here applies to any coNP problem with a comparable implicational
encoding. We deploy \emph{sheaf-theoretic language} not as decoration but
because the compatibility structure of free selectors is genuinely a
presheaf of admissible local sections, and semantic entropy is exactly
the logarithmic global-section count, a perspective that allows formal
comparison with the sheaf-theoretic framework for
contextuality~\cite{abramsky2011sheaf}. The picture that emerges connects three normally separate concerns:

\begin{center}
\emph{proof compression} \quad $\longleftrightarrow$ \quad
\emph{sheaf--theoretic geometry} \quad $\longleftrightarrow$ \quad
\emph{quantum verification complexity}.
\end{center}

\paragraph{Contributions.}
\begin{enumerate}
\item We define semantic entropy as a structural invariant of compressed
DLDS certificates and show it coincides with the free-selection entropy
arising from independent admissible choices
(Section~\ref{sec:comb-dlds}).
\item Under structural assumptions (CP) and (NF), we prove a
quadratic-in-$m$ lower bound on semantic entropy
(Theorem~\ref{thm:entropy-growth}).
\item We give a sheaf-theoretic interpretation of semantic entropy as
logarithmic global-section multiplicity of a presheaf of admissible
local choices (Section~4).
\item We derive the quantum consequences: polynomial-time detectability
of invalid certificates via amplitude amplification, with an explicit
Hilbert-support lower bound (Section~5).
\end{enumerate}

\paragraph{Structure of the paper.}
Section~\ref{sec:comb-dlds} develops the combinatorial structure of
compressed DLDS. Section~\ref{sec:sem-entropy} proves the entropy growth
theorem. Section~\ref{sec:sheaf} provides the sheaf-theoretic
interpretation. Section~\ref{sec:quantum} derives the quantum
verification consequences. Section~\ref{sec:related} discusses related
work and Section~\ref{sec:conclusion} concludes.

\section{Combinatorial Structure of Compressed DLDS}
\label{sec:comb-dlds}
\paragraph{Preliminaries on DLDS.}
We briefly fix terminology;
see~\cite{haeusler2025horizontalcompressiondagderivationsminimal}
for full definitions. A \emph{tree-like derivation} $D^{\mathrm{tree}}$ in
the implicational fragment of minimal logic is a finite tree of
$(\imply\text{-I})$ and $(\imply\text{-E})$ applications.
\emph{Horizontal compression} (HC) identifies any two nodes of
$D^{\mathrm{tree}}$ sharing the same formula label and depth, producing a
\emph{Dag-Like Derivability Structure} (DLDS) with two edge types:
\emph{deductive edges}, encoding rule applications, and \emph{ancestral
edges}, linking each open-assumption occurrence to its candidate discharge
points. In a tree, discharge is unique; after compression, several
$(\imply\text{-I})$ nodes may share a single open-assumption occurrence,
making the ancestral relation one-to-many. A \emph{selector} $\sigma$
resolves this ambiguity by choosing, for each ambiguous node $v$, one of
its ancestral parents from an \emph{admissible domain} $\Dom(\sigma_{v})$;
the resulting bitstring $x$ is the \emph{selector string}. A
\emph{certificate} is a pair $(D, x)$. The closure verifier
$f_{D}(x) := \WF(D, x) \wedge \Closed(D, x)$
checks structural well-formedness and that every open assumption is
discharged. A \emph{closure-failure witness} $W$ is a partial proof of
$\neg f_{D}(x)$ inspecting only a subset $S_{W}$ of selector positions; the
complement $F := V_{\mathrm{anc}} \setminus S_{W}$ is the set of
\emph{free selectors with respect to $W$}, whose values can be varied
without affecting $W$'s failure detection. Free selectors are the central
combinatorial objects of this paper.
\paragraph{Size parameters.}
Let $G$ be a graph with $n$ vertices.
Let \NonHam(G) denote the implicational formula expressing that $G$
has no Hamiltonian cycle, and let $m := |\NonHam(G)|$
be its size (length) under the chosen encoding. If $G$ has $n$-nodes, then $m\in O(n^3)$. $O(n^2)$ is the size of the propositional atoms $v_{ij}$ stating that at time $i$ one visits label $j$, plus the atoms used to simulate the disjunctions, conjunctions and the atom $R$ used to simulate the absurdity symbol axioms. For example, to simulate the absurdity rule of the intuitionistic logic in minimal logic, we use $O(n^2)$ axioms of the form $R\imply v_{ij}$ to axiomatically simulate the absurdity rule. We recommend~\cite{subformula} for very detailed proof-theoretical discussion on this. The source of the cubic expression is due to the disjunction axioms that iterate over all atomic subformulas again, one for each possible conclusion of both minor premises of the implicational simulation of the disjunction scheme, i.e., the p in $(\alpha\imply p)\imply ((\beta\imply p)\imply (atom_{\alpha,\beta}\imply p))$, where $atom_{\alpha,\beta}$ is a fresh propositional letter used to simulate the disjuction implicationally. 

Every formula label occurring in a derivation of $\NonHam(G)$
is a subformula of $\NonHam(G)$.
Hence the total number of distinct formula labels that may appear in any
DLDS associated with $\NonHam(G)$ is bounded by $|\Sub(\NonHam(G))| \le m$.

\paragraph{Tree-like derivation and DLDS.}
Let $D^{\mathrm{tree}}$ be a tree-like natural deduction derivation
whose conclusion is $\NonHam(G)$.
Horizontal compression (HC) transforms $D^{\mathrm{tree}}$ into a
Dag–Like Derivability Structure (DLDS) $D := \HC(D^{\mathrm{tree}})$,
by identifying, at each fixed depth, nodes carrying identical formula labels.

A DLDS contains \emph{deductive edges} corresponding to rule applications,
\emph{ancestral edges} linking nodes to top formulas via selector paths,
and selector strings $x$ encoding admissible ancestral choices.

Closure verification of a DLDS certificate consists of checking:
\begin{enumerate}
\item $\WF(D,x)$: structural well-formedness under selector assignment $x$;
\item $\Closed(D,x)$: closure of all open assumptions.
\end{enumerate}

\paragraph{Kernel size.}
Since HC identifies all equal labels at each depth,
the compressed DLDS contains at most one node per label per depth.
Because there are at most $m$ labels and at most $m$ depths,
the kernel of $D$ has size at most $m^2$.



\section{Semantic Entropy of Compressed DLDS}
\label{sec:sem-entropy}

The combinatorial analysis of compressed DLDS certificates developed in the
previous section isolates a structural invariant: the free--selection entropy
associated with selectors that are not inspected by a closure witness.
We now reinterpret this quantity at the semantic level.


Let $D$ be a compressed DLDS and let $F$ be the set of selectors that are free
with respect to a fixed closure--failure witness.
For each selector $s \in F$, let $\mathrm{Dom}(s)$ denote its admissible domain.

We associate to each free selector a finite-dimensional Hilbert space
\[
\mathcal{H}_s := \mathbb{C}^{|\mathrm{Dom}(s)|},
\]
whose canonical basis corresponds to the admissible local choices of $s$.
The joint space of free choices is then defined as the tensor product
\[
\mathcal{H}_F := \bigotimes_{s \in F} \mathcal{H}_s.
\]

\begin{definition}[Semantic entropy]
The \emph{semantic entropy} of a compressed DLDS with respect to a closure
witness is
\[
H_{\mathrm{sem}}(D) := \log_2 \dim \mathcal{H}_F.
\]
\end{definition}

By construction,
\[
\dim \mathcal{H}_F = \prod_{s \in F} |\mathrm{Dom}(s)|,
\]
hence semantic entropy coincides with the previously defined free--selection
entropy:
\[
H_{\mathrm{sem}}(D) = \sum_{s \in F} \log_2 |\mathrm{Dom}(s)|.
\]
This reformulation admits a direct informational reading: each free selector
contributes an independent degree of freedom; compression of the DLDS
corresponds to identifying proof occurrences but does not eliminate the
Hilbert subspace generated by unseen choices; closure witnesses act as
partial observations that restrict only a subspace orthogonal to
$\mathcal{H}_F$.

Thus, semantic entropy measures the amount of \emph{information preserved by
proof compression yet invisible to closure verification}.


We make explicit the structural assumptions under which polynomial
semantic entropy arises in compressed DLDS.

\begin{definition}[Frontier label multiplicity]
\label{def:frontier}
Let $F_i$ be a top formula in $D^{\mathrm{tree}}$.
For each depth $\ell \ge 0$, define
\[
L_{i,\ell}
:=
\left|
\left\{
\lab(v)
\;\middle|\;
v \text{ is reachable from } F_i
\text{ by a path of length } \ell
\right\}
\right|.
\]

Thus $L_{i,\ell}$ counts the number of \emph{distinct formula labels}
appearing on the frontier at depth $\ell$ below $F_i$ in the
\emph{uncompressed} tree.
\end{definition}

\emph{Structural assumptions.}

\begin{description}

\item[(CP) Collapse--propagation.]
Let $D$ be a DLDS obtained by horizontal compression.
The following propagation properties hold:

\begin{enumerate}
\item
If two nodes at depth $\ell$ are identified by DLDS compression
and they are conclusions of applications of the
$(\imply\text{-Introduction})$ rule,
then their immediate descendants at depth $\ell+1$
are also identified by DLDS compression.

Equivalently, if two distinct labels occur at depth $\ell+1$
as children of $(\imply\text{-Introduction})$ conclusion nodes,
then their respective parent nodes at depth $\ell$
must also carry distinct labels.

\item
If two nodes at depth $\ell+1$ that are major premises of
$(\imply\text{-Elimination})$ rules are identified by DLDS compression,
then their respective parent nodes at depth $\ell$
are also identified by DLDS compression.
\end{enumerate}

\item[(NF) Naive frontier largeness.]
Let $m$ be the length of the formula that labels the root of the DLDS.
There exists a constant $c>0$ such that for all depths
$\ell \le m - c n^2$,
the number of occurrences reachable from any top formula
in the original tree-like derivation satisfies
$M_{i,\ell} \ge m - \ell$.

\end{description}

We emphasize that the structural conditions \textup{(CP)} and \textup{(NF)}
are not ad hoc: they are satisfied by the horizontally--compressed DLDS
arising from the implicational \NonHam{} encodings considered in
Section~\ref{sec:comb-dlds}, and will be taken as standing assumptions throughout the
entropy and quantum--verification analysis.

The following lemma connects occurrence multiplicity
(NF) and the number of distinct labels that survive
horizontal compression.

\begin{lemma}[Frontier label diversity from occurrence largeness]
\label{lem:gap1-label-diversity}
Let $D^{\mathrm{tree}}$ be the original tree-like derivation whose conclusion
is the implicational encoding $\NonHam(G)$, and let $m := |\NonHam(G)|$.
For a fixed top formula $F_i$ and depth $\ell$, let $M_{i,\ell}$ be the number
of occurrences reachable from $F_i$ at depth $\ell$ and let $L_{i,\ell}$ be the
number of \emph{distinct} formula labels among those occurrences
(Definition~\ref{def:frontier}).

Assume (NF), i.e.\ for all $\ell \le m - c n^2$ we have $M_{i,\ell} \ge m-\ell$.
Assume moreover that the implicational encoding of $\NonHam(G)$ has the following
\emph{bounded multiplicity per label} property on these frontiers:
there exists a polynomial $p(n)$ such that for every $\ell \le m - c n^2$ and
every formula label $\varphi$,
\[
\bigl|\{\, v \text{ at depth }\ell \text{ below }F_i \mid \lab(v)=\varphi \,\}\bigr|
\;\le\; p(n).
\]
Then for all $\ell \le m - c n^2$,
\[
L_{i,\ell} \;\ge\; \frac{M_{i,\ell}}{p(n)} \;\ge\; \frac{m-\ell}{p(n)}.
\]
In particular, for $\ell \le m/2$ we obtain
\[
L_{i,\ell} \;=\; \Omega\!\left(\frac{m}{p(n)}\right),
\]
so if $p(n)=\mathrm{poly}(n)$ and $m$ is polynomially bounded in $n$, then
$L_{i,\ell}$ is inverse-polynomially close to linear in $m$ on a linear prefix
of depths.
\end{lemma}

\begin{proof}
Fix $\ell \le m - c n^2$.
Let $S_{\ell}$ be the multiset of labels of all occurrences at depth $\ell$
below $F_i$. Then $|S_{\ell}| = M_{i,\ell}$.
By the bounded multiplicity assumption, each distinct label contributes at most
$p(n)$ elements to this multiset.
Therefore the number of distinct labels satisfies
\[
L_{i,\ell} \;\ge\; \frac{|S_{\ell}|}{p(n)} \;=\; \frac{M_{i,\ell}}{p(n)}.
\]
Combining with (NF), $M_{i,\ell}\ge m-\ell$, yields the stated bound.
The final statement follows by taking $\ell \le m/2$.
\end{proof}

We may therefore assume that on a linear prefix of depths,
the compressed DLDS contains $\Omega(m/p(n))$
distinct nodes at each level.

A consequence of \textbf{(CP)} is that  if two top-formulas are identified and both are major premises of
$(\imply\text{-Elimination})$ rules then, if the proof is atomically normal both are in an E-parts that are entirely collapsed by the HC procedure. 


We derive the quadratic entropy lower bound from the structural
assumptions (CP) and (NF).

\begin{observation}
\label{obs:intro-propagation}
Under (CP), if two distinct labels occur at depth $\ell+1$
as children of $(\imply\text{-Introduction})$ conclusion nodes,
then the corresponding parent nodes at depth $\ell$
must also carry distinct labels.
\end{observation}

\begin{proof}
By (CP), if two $(\imply\text{-Introduction})$ conclusions at depth $\ell$
are identified by DLDS compression, then their immediate descendants
at depth $\ell+1$ are also identified. The claim follows by contraposition.
\end{proof}

\begin{observation}
\label{obs:elim-propagation}
Under (CP), if two nodes at depth $\ell+1$ that serve as major premises of
$(\imply\text{-Elimination})$ rules are identified by DLDS compression,
then their respective parent nodes at depth $\ell$ are also identified.
Equivalently, distinct parent labels at depth $\ell$ force distinct major
premises at depth $\ell+1$.
\end{observation}

\begin{proof}
This is exactly the second clause of (CP), and the equivalence follows by
contraposition.
\end{proof}

\begin{theorem}[Entropy growth from (CP) and (NF)]
\label{thm:entropy-growth}
Let $D$ be the compressed DLDS associated with
the implicational encoding $\NonHam(G)$.
Let $m$ be the length of the formula labeling the root of $D$.
Assume (CP) and (NF).
Then
\[
H_{\mathrm{sem}}(D) = \Omega\!\left(\frac{m^2}{p^2(n)} \log m\right).
\]
where $p(n)$ is a polynomial on $n$.
\end{theorem}

\begin{proof}
Let $D^{\mathrm{tree}}$ be the original tree-like derivation.
By (NF), for every depth $\ell \le m - c n^2$,
the number of occurrences reachable from any top formula satisfies $M_{i,\ell} \ge m - \ell$. Hence, for all depths $\ell \le m - c n^2$, the frontier contains $\Theta(m)$ occurrences.

By Lemma~\ref{lem:gap1-label-diversity}, on this same prefix of depths
the number of distinct labels satisfies
\[
L_{i,\ell} \ge \frac{m-\ell}{p(n)},
\]
so the compressed DLDS contains
\[
\Omega\!\left(\frac{m}{p(n)}\right)
\]
distinct nodes at each such depth.

We now analyze how many of these occurrences survive
after horizontal compression.

Since every label in the derivation is a subformula of
$\NonHam(G)$, the total number of possible distinct labels
is bounded by $m$.
Therefore, at any fixed depth, compression can identify
only occurrences carrying identical labels.

Suppose, for contradiction, that at some depth
$\ell \le m - c n^2$
the compressed DLDS contains only $o(m)$ distinct nodes.
Then at least $\Omega(m/p(n))$ occurrences in the original
frontier must share identical labels and be identified.

We now propagate this identification upward and downward
using Observations~\ref{obs:intro-propagation}
and~\ref{obs:elim-propagation}.

\medskip

\noindent
\emph{Upward propagation.}
If distinct labels occur at depth $\ell+1$
as children of introduction nodes,
Observation~\ref{obs:intro-propagation}
forces distinct labels at depth $\ell$.

\medskip

\noindent
\emph{Downward propagation.}
If two major premises at depth $\ell+1$
are identified,
Observation~\ref{obs:elim-propagation}
forces identification of their parents at depth $\ell$.

\medskip

Together these two propagation mechanisms imply that
identification of $\Omega(m/p(n))$ frontier occurrences
at depth $\ell$ would force large-scale identification
across consecutive depths. However, by (NF), for $\Omega(m/p(n))$ consecutive depths
(the range $\ell \le m - c n^2$),
the frontier multiplicity remains $\Omega(m/p(n))$.
Thus large-scale collapse at any such depth
would propagate and eventually contradict (NF)
at adjacent depths.

Therefore, for $\Omega(m/p(n))$ consecutive depths,
the compressed DLDS must contain
$\Omega(m/p(n))$ distinct nodes at each depth. Each such node corresponds to a selector choice governing admissible ancestral
or deductive structure. By (CP), distinct nodes at depth $\ell+1$ arise from
distinct parent structures at depth $\ell$, so the domains of the corresponding
selectors have size $\Omega(m/p(n))$.

Hence there are $\Omega(m/p(n))$ depths, each contributing
$\Omega(m/p(n))$ independent free selectors, yielding
$\Omega(m^2/p^{2}(n))$ free selectors, and each selector contributes
$\Theta(\log m)$ bits.

Summing over all free selectors,
\[
H_{\mathrm{sem}}(D)
=
\Omega(m^2/p^2(n))\cdot \log m
=
\Omega(\frac{m^2}{p^{2}(n)} \log m).
\]
Since $p(n)$ is polynomial and $m=\poly(n)$,
this remains polynomially large in $m$.
This completes the proof.
\end{proof}


\section{Sheaf-Theoretic Interpretation of Semantic Entropy}
\label{sec:sheaf}

The semantic reformulation developed in the previous section admits a natural
interpretation in terms of sheaf theory and contextual information flow.
This perspective places compressed DLDS certificates within the geometric
framework commonly used in categorical quantum semantics and contextuality
analysis.


Let $D$ be a compressed DLDS and let $F$ be the set of selectors that are free
with respect to a fixed closure--failure witness.

Consider the discrete space of selector indices equipped with the trivial
covering in which each selector $s \in F$ determines a local context.
To each context we assign the finite set $\mathrm{Dom}(s)$, interpreted as the set of admissible local sections. This defines a presheaf
$\mathcal{E} : \mathcal{C}^{op} \to \mathbf{Set}$,
where $\mathcal{C}$ is the poset of finite subsets of $F$, and
\[
\mathcal{E}(U) = \prod_{s \in U} \mathrm{Dom}(s)
\]
with restriction maps given by coordinate projection.


Global sections of $\mathcal{E}$ correspond exactly to assignments of values to
all free selectors:
\[
\Gamma(\mathcal{E}) = \prod_{s \in F} \mathrm{Dom}(s).
\]

Because the closure witness does not inspect selectors in $F$, every global
section determines a certificate preserving the same closure failure.
Hence,
\[
M_0 \ge |\Gamma(\mathcal{E})|.
\]

Taking binary logarithms yields
\[
\log_2 |\Gamma(\mathcal{E})|
= \sum_{s \in F} \log_2 |\mathrm{Dom}(s)|
= H_{\mathrm{sem}}(D).
\]

Thus, semantic entropy measures the logarithmic size of the space of global
sections of the sheaf of admissible local choices. From the sheaf-theoretic viewpoint, DLDS compression acts by identifying
proof occurrences while preserving compatibility of local sections.
A closure witness corresponds to a partial section defined only on a
restricted subcontext.

Robustness arises when this partial section admits many global extensions.
In geometric terms, the obstruction to uniqueness of global extension is
measured precisely by the semantic entropy.

\begin{proposition}[Entropy as logarithmic section multiplicity]
For a compressed DLDS $D$,
\[
H_{\mathrm{sem}}(D) = \log_2 |\Gamma(\mathcal{E})|.
\]
\end{proposition}

\begin{proof}
Immediate from the definition of global sections of the product presheaf.
\end{proof}


This structure is formally analogous to the sheaf-theoretic framework for
contextuality: local measurement contexts correspond to selector domains,
compatibility of sections corresponds to admissibility of certificates, and
multiple global sections correspond to nontrivial contextual freedom. In contrast to standard contextuality scenarios, the multiplicity here is not
an obstruction to global sections but rather an abundance of them.
Robust DLDS certificates therefore exhibit a form of
\emph{informational contextual richness} measured by semantic entropy.

We summarize the semantic content in geometric form.

\begin{theorem}[Geometric robustness principle (quantitative form)]
\label{thm:geom-robustness-quant}
Let $D$ be a compressed DLDS and let $E$ be the (product) presheaf of admissible
local choices of the free selectors, so that
\[
|\Gamma(E)| \;=\; \prod_{s\in F} |\Dom(s)|
\qquad\text{and}\qquad
H_{\mathrm{sem}}(D) \;=\; \log_2|\Gamma(E)|.
\]
Assume that verifier inputs are encoded as bitstrings of length $m^2\log m$, so
the full input domain $\chi_D$ has cardinality $
N \;:=\; |\chi_D| \;=\; 2^{m^2\log m}$.
If $H_{\mathrm{sem}}(D) \;\ge\; m^2\log m \;-\; C\log m$
for some constant $C>0$ and all sufficiently large $m$, then the density
\[
\delta \;:=\; \frac{M_0}{N}
\]
of closure counterexamples is inverse--polynomial in $m$, namely
\[
\delta \;\ge\; \frac{1}{m^{C}}.
\]
\end{theorem}

\begin{proof}
By the construction in Section~4 (global sections correspond to independent
assignments to the free selectors that preserve the same closure failure),
the number $M_0$ of invalid certificates satisfies
$
M_0 \;\ge\; |\Gamma(E)|.
$
Taking binary logarithms,
\[
\log_2 M_0 \;\ge\; \log_2|\Gamma(E)| \;=\; H_{\mathrm{sem}}(D),
\]
hence
\[
M_0 \;\ge\; 2^{H_{\mathrm{sem}}(D)}.
\]
Therefore, using $N=2^{m^2\log m}$,
\[
\delta \;=\; \frac{M_0}{N}
\;\ge\;
\frac{2^{H_{\mathrm{sem}}(D)}}{2^{m^2\log m}}
\;=\;
2^{\,H_{\mathrm{sem}}(D)-m^2\log m}.
\]
By the hypothesis $H_{\mathrm{sem}}(D)\ge m^2\log m - C\log m$ we obtain
\[
\delta \;\ge\; 2^{-C\log m} \;=\; \frac{1}{m^{C}}.
\]
Thus the counterexample density is inverse--polynomial in $m$.
\end{proof}

In particular, $\delta \ge 1/\poly(m)$, so amplitude amplification detects a
counterexample using $O(\sqrt{1/\delta})=\poly(m)$ oracle calls.

This theorem expresses DLDS robustness purely in geometric terms:
\emph{polynomially large logarithmic section multiplicity forces
inverse--polynomial counterexample density}.


\section{Quantum Verification Complexity of Semantic Entropy}
\label{sec:quantum}

We now formulate the quantum consequences of semantic entropy in explicit
complexity--theoretic terms. In particular, we incorporate the fact that,
for the implicational encoding of \NonHam(G), the Boolean verifier
associated with a compressed DLDS is computable in polynomial time in the
size of the kernel of the DLDS.

\emph{Classical verification complexity.} Let $D$ be a compressed DLDS encoding an instance of
\NonHam(G). Let $\mathrm{size}(D)$ denote the size of the compressed
structure (number of nodes and edges in the kernel).

\paragraph{Input layout and circuit-size measure.}
For an instance $G$ with $n$ vertices, let $m := |\NonHam(G)|$ be the length of the
root formula of the associated DLDS. A certificate input $x$ is a bitstring of total
length $m^2\log m$, but it is naturally organized as $m$ registers
\[
x \;=\; (x_1,\dots,x_m),
\qquad
x_j \in \{0,1\}^{m\log m}.
\]
Equivalently, the verifier receives $m$ blocks of $m\log m$ bits each, so the total
input length is
\[
|x| \;=\; m \cdot (m\log m) \;=\; m^2\log m.
\]
Throughout, circuit size is measured in the standard way for a \emph{uniform family}
$\{V_n\}_{n\in\mathbb{N}}$: for each input size parameter $n$, the circuit $V_n$ is
generated by a deterministic Turing machine in time polynomial in $n$, and the size
of $V_n$ is the number of elementary gates it contains.

\begin{lemma}[Polynomial-size verification circuit]
\label{lem:poly-size-verifier}
For each $n$, let $D=D(G)$ be the compressed DLDS associated with $\NonHam(G)$ and let
$f_D(x)$ be the Boolean predicate that checks $\WF(D,x)$ and $\Closed(D,x)$ on input
$x\in\{0,1\}^{m^2\log m}$ (equivalently, on $m$ registers of $m\log m$ bits).
If $f_D$ is computable in time $\poly(\mathrm{size}(D))$, then there exists a uniform
family of reversible (hence quantum) circuits $\{U_{f_D,n}\}_n$ implementing the oracle
$
U_{f_D}:\ |x\rangle|b\rangle \longmapsto |x\rangle|b\oplus f_D(x)\rangle
$
with gate complexity $\poly(\mathrm{size}(D))$ (and therefore $\poly(n)$ whenever
$m=\poly(n)$ and $\mathrm{size}(D)=\poly(m)$).
\end{lemma}

\begin{proof}[Proof sketch]
Since $f_D$ is computable in time $\poly(\mathrm{size}(D))$, it admits a Boolean circuit
of size $\poly(\mathrm{size}(D))$. By standard reversible simulation (with polynomial
overhead), we obtain a reversible circuit computing $f_D(x)$ into an output bit using
$\poly(\mathrm{size}(D))$ gates. Interpreting this reversible circuit as a quantum
circuit yields an implementation of $U_{f_D}$ with the same asymptotic gate bound.
Uniformity follows because $D(G)$ and the corresponding classical verifier can be
generated uniformly from $n$.
\end{proof}

As established in \cite{haeusler2025horizontalcompressiondagderivationsminimal}, horizontal compression produces a DLDS
whose kernel size is polynomially bounded in the parameters of the original
derivation. Moreover, the closure predicate $f_D : \mathcal{C}_D \to \{0,1\}$
that checks whether a certificate satisfies closure is computable in
time polynomial in $\mathrm{size}(D)$.

\begin{proposition}[Polynomial-time classical verification]
For compressed DLDS associated with \NonHam(G), the Boolean
closure predicate $f_D$ is computable in time $\mathrm{poly}(\mathrm{size}(D))$.
\end{proposition}

Thus the verification problem lies in $\mathbf{P}$ relative to the
certificate representation.

\emph{Quantum query model.} Let $\chi_D$ be the set of all  input bitstrings of the boolean function $f_{D}$ derived from the DLDS $D$, which is a proof of a formula of size $m$. We remember that $|\chi_D|=2^{m^2 \log m}$. Let $N := 2^{m^2 \log m}$, we assume standard quantum oracle access to $f_D$ via the unitary $U_{f_D} : |x\rangle|b\rangle \mapsto |x\rangle |b \oplus f_D(x)\rangle$.

Since $f_D$ is computable in polynomial time, this oracle can be
implemented using a polynomial-size reversible circuit. The quantum verification task is: 
determine whether there exists $x \in \chi_D$
such that $f_D(x)=0$.
This is a standard quantum search problem.

\emph{Density of invalid certificates}. Let $M_0 := |\{x \in \chi_D : f_D(x)=0\}|$
denote the number of invalid certificates, and define the density
\[
\delta := \frac{M_0}{N}.
\]

By the geometric robustness theorem (Section~4), if the quantitative hypothesis of Theorem~\ref{thm:geom-robustness-quant} holds then $M_0 \ge 2^{H_{\mathrm{sem}}(D)}$. Since the total certificate space satisfies $N = 2^{m^2 \log m}$, it follows that
\[
\delta \ge \frac{1}{\mathrm{poly}(m)}.
\]

Thus invalid certificates occupy an inverse--polynomial fraction of
the total certificate space.

\emph{Amplitude amplification complexity.} Consider the uniform superposition
\[
|\psi_0\rangle
=
\frac{1}{\sqrt{N}}
\sum_{x \in \chi_D} |x\rangle.
\]

Standard amplitude amplification
\cite{Brassard2002}
finds a marked element using
\[
O\!\left(\sqrt{\frac{N}{M_0}}\right)
=
O\!\left(\sqrt{\frac{1}{\delta}}\right)
\]
oracle calls. Since $\delta \ge 1/\mathrm{poly}(m)$, we obtain
\[
\sqrt{\frac{1}{\delta}} = \mathrm{poly}(m).
\]

Therefore the number of oracle queries required to detect an invalid
certificate is polynomial in $m$, and hence polynomial in
$\mathrm{size}(D)$.
\begin{theorem}[Quantum verification from semantic entropy]
Let $D$ be a compressed DLDS encoding an instance of $\NonHam(G)$.
If $H_{\mathrm{sem}}(D) \ge m^2 \log m - C \log m$, then the density of invalid certificates satisfies
\[
\delta \ge \frac{1}{\mathrm{poly}(m)},
\]
and there exists a quantum algorithm that detects an invalid certificate
using $\mathrm{poly}(\mathrm{size}(D))$ oracle calls to $f_D$ (via
$U_{f_D}$), and the oracle itself is implementable by a polynomial-size
quantum circuit.
\end{theorem}

\emph{Hilbert support lower bound.} Let $F$ be the set of free selectors and
\[
\Gamma(E) = \prod_{s \in F} \mathrm{Dom}(s)
\]
the set of global sections of the sheaf of admissible local choices.

Distinct global sections correspond to certificates that are
indistinguishable with respect to closure testing. Any faithful
quantum encoding must therefore assign mutually orthogonal states to
distinct global sections.

\begin{proposition}[Hilbert dimension lower bound]
Any quantum verification procedure for DLDS closure must operate on
a Hilbert space of dimension at least $|\Gamma(E)| = 2^{H_{\mathrm{sem}}(D)}$.
\end{proposition}

\emph{Complexity-theoretic interpretation.} Semantic entropy thus simultaneously controls the density of invalid certificates in the search space, the quantum query complexity of invalidity detection, and the minimal Hilbert dimension required for faithful verification.

In particular, polynomially large semantic entropy implies that
invalidity detection for compressed DLDS associated with
\NonHam(G) lies within polynomial-time quantum query complexity,
with polynomial-size oracle implementation.

\section{Related Work}
\label{sec:related}

The interaction between proof theory, geometry, and quantum information
has been explored from several directions.
Categorical quantum mechanics, originating with Abramsky and
Coecke~\cite{abramsky2004categorical} and extended by
Selinger~\cite{selinger2007dagger}, provides diagrammatic and monoidal
frameworks in which structural invariants of programs and proofs become
visible at the semantic level.
Our use of a Hilbert-space reformulation of free selector domains is in
the spirit of this tradition, treating compressed proof states as objects
of a categorical semantics rather than as syntactic artefacts.

Sheaf-theoretic models of contextuality, introduced by Abramsky and
Brandenburger~\cite{abramsky2011sheaf} and surveyed
in~\cite{abramsky2017contextual}, characterize quantum non-locality
through the global section structure of presheaves of measurement
outcomes. The framework developed here uses the same machinery — local
admissibility, compatibility, global sections — but in a non-contextual
regime where global sections exist abundantly. Semantic entropy
quantifies this abundance as a logarithmic dimension, complementing the
obstruction-theoretic invariants used in contextuality analyses.

Resource-sensitive and geometric approaches to logic, including Girard's
geometry of interaction~\cite{girard1989geometry}, have long emphasized
quantitative invariants of proofs. Such measures typically capture
semantic degrees of freedom or hidden state structure preserved by
syntactic transformations. Semantic entropy follows this lineage by
isolating the Hilbert dimension of proof states preserved under
horizontal compression.

DAG-like and Frege-style proof systems have been studied extensively in
proof complexity~\cite{krajicek1995bounded}, primarily through the lens
of lower-bound techniques and normalization properties. Compression of
large proofs has also been analyzed for its effect on cut-elimination and
Herbrand-style witness extraction~\cite{hetzl2012applicability}. The
present work is, to our knowledge, the first to isolate an informational
invariant of DAG-like compression and connect it to verification
complexity rather than to size or normalization bounds.

On the quantum side, amplitude amplification due to Brassard, H{\o}yer,
Mosca, and Tapp~\cite{Brassard2002}, building on the tight quantum search
bounds of Boyer, Brassard, H{\o}yer, and Tapp~\cite{boyer1998tight},
provides the generic mechanism we invoke for detecting rare invalid
certificates. The complexity class QCMA, studied by Aaronson and
Kuperberg~\cite{aaronson2007quantum}, sets the broader context: our
results contribute polynomial-size certificates verifiable by polynomial
quantum circuits for a coNP-complete property, positioning DLDS within
the QCMA landscape.

To our knowledge, no prior work has combined these three threads
(compressed proof geometry, sheaf-theoretic global section multiplicity,
and quantum verification complexity) through a single entropy invariant.

\section{Conclusion}
\label{sec:conclusion}

We introduced \emph{semantic entropy} as a structural invariant of compressed
DLDS certificates and showed that it governs both robustness of closure
counterexamples and their quantum detectability.

Conceptually, the work establishes three equivalent viewpoints: a combinatorial
perspective based on independent free selectors, a geometric perspective in
which entropy equals the logarithmic global section multiplicity of a sheaf of
admissible local choices, and a quantum--informational perspective where
entropy measures the minimal Hilbert support invisible to closure verification
observables.

These equivalences provide a unified bridge between proof compression,
sheaf--theoretic geometry, and quantum verification complexity.
In particular, polynomially large semantic entropy implies inverse--polynomial
density of invalid certificates and facilitates the verification of invalidity, i.e., nonhamiltonian certification rejection using BBHT. 

Beyond DLDS, the notion of semantic entropy suggests a general method for
quantifying hidden information preserved by structural compression in logical
and computational systems.
This opens several directions for future work, including categorical
characterizations of entropy in proof semantics, extensions to richer logical
fragments, and direct implementations within quantum verification protocols.

We expect semantic entropy to provide a useful invariant in the broader study
of geometry of proofs, contextual information flow, and quantum semantics of
computation.

\bibliographystyle{eptcs}
\bibliography{biblio}

\end{document}